# SDU
University of Southern Denmark

Business Models for Digitalization Enabled Energy Efficiency and Flexibility in Industry: A Survey with Nine Case Studies

Ma, Zhipeng; Jørgensen, Bo Nørregaard; Levesque, Michelle; Amazouz, Mouloud; Ma, Zheng Grace

Go to publication entry in University of Southern Denmark's Research Portal

# Business Models for Digitalization Enabled Energy Efficiency and Flexibility in Industry: A Survey with Nine Case Studies

Zhipeng Ma[1]([✉]) [0000-0002-4049-539X], Bo Nørregaard Jørgensen[1]([✉]) [0000-0001-5678-6602], Michelle Levesque[2], Mouloud Amazouz[3], Zheng Grace Ma[1]([✉]) [0000-0002-9134-1032]

[1] SDU Center for Energy Informatics, Maersk Mc-Kinney Moller Institute, The Faculty of Engineering, University of Southern Denmark, 5230 Odense, Denmark
{zhma, bnj, zma}@mmmi.sdu.dk
[2] CanmetMINING-Sudbury, Natural Resources Canada, 1079 Kelly Lake Road Sudbury, ON P3E 5P5 Canada
michelle.levesque@NRCan-RNCan.gc.ca
[3] CanmetENERGY Varennes, Natural Resources Canada, 1615 Lionel-Boulet Blvd. Varennes, QC J3X 1S6 Canada
mouloud.amazouz@NRCan-RNCan.gc.ca

**Abstract.** Digitalization is challenging in heavy industrial sectors, and many pilot projects facing difficulties to be replicated and scaled. Case studies are strong pedagogical vehicles for learning and sharing experience & knowledge, but rarely available in the literature. Therefore, this paper conducts a survey to gather a diverse set of nine industry cases, which are subsequently subjected to analysis using the business model canvas (BMC). The cases are summarized and compared based on nine BMC components, and a Value of Business Model (VBM) evaluation index is proposed to assess the business potential of industrial digital solutions. The results show that the main partners are industry stakeholders, IT companies and academic institutes. Their key activities for digital solutions include big-data analysis, machine learning algorithms, digital twins, and Internet of Things developments. The value propositions of most cases are improving energy efficiency and enabling energy flexibility. Moreover, the technology readiness levels of six industrial digital solutions are under level 7, indicating that they need further validation in real-world environments. Building upon these insights, this paper proposes six recommendations for future industrial digital solution development: fostering cross-sector collaboration, prioritizing comprehensive testing and validation, extending value propositions, enhancing product adaptability, providing user-friendly platforms, and adopting transparent recommendations.

# 1    Introduction

The industry sector accounts for 25.6% of the energy consumption in the European Union (EU) in 2021 [1], where both the industrial process and non-process-related operations contribute to this energy usage. It is crucial to improve energy efficiency and flexibility in the industry to combat global climate change. In this context, digitalization has emerged as a promising technology to achieve these goals in Industry 4.0 [2].

Energy efficiency [3] refers to the capacity of decreasing energy consumption while maintaining production quality. It involves employing technologies, practices, and systems that minimize energy waste and maximize the output obtained from the energy consumed [4]. On the other hand, energy flexibility [5] pertains to the capability of an energy system to adapt its energy production or consumption patterns in response to fluctuations in energy supply or demand conditions. This flexibility allows the system to efficiently balance the energy generated or consumed with the changing needs and availability of energy resources [6].

Digitalization enables the collection and analysis of large-scale data from industrial processes, equipment, and operations for the research of energy efficiency and flexibility through various technologies such as Artificial Intelligence (AI), Machine Learning (ML), Internet of Things (IoT), Digital Twins (DT), Cloud Computing, etc [7]. These technologies can be leveraged to identify energy consumption patterns [8] and optimize energy usage [9] in industry. Moreover, industrial processes can be optimized to minimize energy usage and maximize production quality with Optimization models and ML algorithms [10]. They can be employed to identify energy-efficient operating conditions [11], adjust setpoints [3], and optimize process parameters [12]. Furthermore, the intelligent system can offer recommendations for optimizing key devices regarding energy efficiency or flexibility by utilizing explainable AI frameworks[13]. These frameworks provide transparency and comprehensibility in the decision-making process, enabling users to understand and interpret the provided recommendations.

Numerous projects have emerged to leverage the benefits of digitalization by employing digital tools to assist industrial stakeholders in attaining their energy consumption goals [14]. However, the process of industrial digitalization is intricate and influenced by various factors [15]. Consequently, developing tailored strategies that cater to distinct requirements and minimize individual risks becomes essential [16].

Prior to the development of industrial digital solutions, it is crucial to conduct a comprehensive survey of existing cases and understand their strengths and drawbacks. The Business Model Canvas (BMC) is a strategic management tool that visually represents key value blocks, outlining the core aspects of a business model [17]. By using the BMC, a systematic overview of industrial digitalization cases can be obtained, enabling individuals to identify potential gaps, opportunities, and areas for innovation within their products, which facilitates more effective decision-making and strategic planning processes. Four business models based on BMC have been developed in [15] to investigate the building participation in energy aggregation market by analyzing building owners' requirements to develop feasible market access strategies for different types of buildings. An evaluation tool [18] is proposed to access the business model for smart cities with the potential to be extend for evaluating all BMCs for digital solutions.

So far, only a few studies have surveyed industrial digitalization cases for energy efficiency and flexibility and employed the BMC to analyze the cases. To fill the research gap, this paper utilizes the BMC to analyze nine cases of digitalization-enabled energy efficiency and flexibility in the industry. Moreover, their values in different BMC blocks are analyzed in detail and an evaluation index is proposed to assess the value of potential business models of industrial digital solutions.

The remainder of the paper is organized as follows. Section 2 reviews the literature on digital solutions and relevant business models for energy efficiency and flexibility. Section 3 introduces the data collection and analysis methods. Section 4 presents the results of the BMC on nine cases in industrial digitalization and section 5 discusses the potentials and challenges of digitalization for energy efficiency and flexibility. Section 6 concludes the study and makes recommendations for future research.

## 2    Related Works

### 2.1    Digital Solutions for Energy Efficiency and Flexibility

Advancements in digitalization technologies, such as Artificial Intelligence (AI), Machine Learning (ML), the Internet of Things (IoT), Digital Twin (DT), and Cloud Computing, have provided unique opportunities for augmenting the performance of energy applications. The fusion of these diverse methodologies often results in a more precise and resilient analysis.

For instance, a study in [19] introduces an innovative ML-based optimization framework to tweak the pivotal parameters affecting bioenergy production, with the objective of enhancing energy efficiency. This study leverages a neural network (NN) for predicting energy consumptions and production yield with a high accuracy.

In another research [8], a kCNN-LSTM framework is put forward to yield accurate predictions on building energy consumption. The research employs k-means clustering to decipher energy consumption patterns; utilizes Convolutional Neural Networks (CNN) to unravel complex features that influence energy consumption; and uses Long Short Term Memory (LSTM) to capture long-term dependencies by modeling temporal information in time series data. This framework's implementation at both the electricity network and user levels could significantly assist in informed decision-making, effective demand management, and improving energy efficiency.

Furthermore, [20] employs a multi-agent-based simulation to probe into the potential of energy flexibility in the Danish brewery sector. The research indicates that Danish breweries could reduce their electricity expenses by 1.56% annually, maintain operational security, and decrease their greenhouse gas (GHG) emissions by approximately 1745 tons. Moreover, [21] applies a life cycle assessment (LCA) model to evaluate the carbon footprint in the food processing industry. Here, the k-means clustering technique is used to discern the optimal production flow that results in lower energy consumption. The study's findings suggest that the most energy-efficient temperature at the end of the production line is -25°C, with a tolerance of ± 2°C.

## 2.2 Business Models for Energy Efficiency and Flexibility

Business models play an important role in digital technology analysis, which is helpful to facilitate, configure, and broker system innovation [22]. [23] discusses the business model innovation to address salient bottlenecks in the European energy transition. The results show that digital technologies for cost reduction are often more impactful than digital technologies for adding value.

The study in [24] indicates that eight dimensions in business models should be considered when analyzing industrial digitalization, including resource, network, customer, market, revenue, manufacturing, procurement, and financial. The summary and comparison demonstrate that digitalization has a positive impact on sustainable manufacturing. Furthermore, [25] introduces a digital business model innovation ladder that helps to enhance understanding of the social and environmental values associated with industrial digitalization. The higher up business models are on the innovation ladder, the greater value they create for the energy system but fewer benefits for users.

Existing business models typically focus on analyzing either digital solutions or energy efficiency studies. However, this study introduces a novel business model that specifically analyzes industrial digital solutions aimed at improving energy efficiency and enhancing energy flexibility. The details of this model will be pre-sented in the following section.

# 3 Methodology

## 3.1 Data Collection

This study is aimed at gathering cases of digitalization adoption and implementation in industries across multiple sectors. To achieve this, a survey methodology is employed, where industrial stakeholders and IT technology providers offer essential information on base details, technology implementation, and other key takeaways. The data collection process involves a questionnaire with open-ended questions to gather the necessary information from the partners.

The questionnaire comprises four sections: basic information, technology development, market segment, and additional information. The basic information section encompasses both the company's and project's brief descriptions. The technology development section gathers information regarding technology applications, such as implemented solutions, partners involved, and associated implementation costs. The market segment section focuses on the anticipated value proposition and potential implementation challenges. Lastly, the additional information section covers key takeaways and lessons learned.

## 3.2 Data Analysis – Business Model Canvas

**Business Model Canvas.** The business model canvas (BMC) [17] is a strategic management tool that provides a visual framework for describing, analyzing, and designing business models, which comprises nine key components, including key partners, key activities, key resources, value propositions, customer relationships, channels, customer segments, cost structure, and revenue streams.

Fig. 1 demonstrates the business model for industrial digitalization for energy efficiency and flexibility based on the nine cases. Academic and industry partners play distinct roles when developing industrial digital solutions and contribute different expertise. Academic partners primarily provide theoretical support and state-of-the-art technology, while industry partners contribute data and assist in tool testing and validation. Value proposition represents the unique values that a system offers to meet the specific needs and demands of its target customers. In the context of energy efficiency and flexibility, the industry focuses on achieving goals such as reducing energy consumption and GHG emissions. Moreover, there is also a need to decrease operational costs and enhance production quality.

In the customer relationship perspective, the system provider should offer customized service after selling their products, because most industrial digital solutions are customized. Hence, both communication channels and revenue streams are also negotiated between providers and customers. Moreover, the customer segments involved can vary significantly due to distinct targets in different cases.

| Key Partners | Key Activities | Value Propositions | Customer Relationships | Customer Segments |
|---|---|---|---|---|
| • Academic stakeholders<br>• Industrial stakeholders | • Big-data analysis<br>• AI | • Energy efficiency<br>• Energy flexibility | • Consultation<br>• Maintenance service | • Production data management<br>• R&D |
| | **Key Resources**<br><br>• Skilled Professionals<br>• Software/Platform | | **Channels**<br><br>• Online platforms<br>• Telephonic service | |
| **Cost Structure**<br><br>• Fix costs<br>• Variable costs | | | **Revenue Streams**<br><br>• Financial inflow | |

**Fig. 1.** Business model of digital solutions for energy efficiency and flexibility in industry.

### 3.3 Evaluation Index for Business Models

An evaluation index is developed to assess the value of potential business models based on the evaluation tools in [15, 18]. All components affecting business models as well as technology readiness levels (TRLs) and Relevance-Breadth (RB) are included in Eqa.(1). This evaluation index can be utilized in all technology-related business model analyses to investigate their value potential. The value 0.3 in Eqa.(1) is aimed at avoiding negative results by shifting the VBM (Value of business model) value. The TRL and RB are set as two independent factors while the remaining parameters are parts in a sum, because the TRL and RB directly relate to commercialization potential whereas the others are support factors in the evaluation [15].

$$VBM = TRL \times RB \times (EPR+RID+RA+NCP+CAD+NS+FA-RIC+0.3) \qquad (1)$$

TRLs are a method for estimating the maturity of technologies during the acquisition phase of a program [26]. In 2013, the TRL scale was further canonized by the International Organization for Standardization (ISO) with the publication of the ISO 16290:2013 standard. Table 1 shows the definition of the TRL scale in the European Union (EU). The TRL level 1-9 is also the value of parameter TRL in Eqa.(1).

**Table 1.** The definition of TRL [27].

| TRL | Definition |
|---|---|
| 1 | Basic principles observed |
| 2 | Technology concept formulated |
| 3 | Experimental proof of concept |
| 4 | Technology validated in the laboratory |
| 5 | Technology validated in a relevant environment |
| 6 | Technology demonstrated in a relevant environment |
| 7 | System prototype demonstration in an operational environment |
| 8 | System complete and qualified |
| 9 | Actual system proven in an operational environment |

The RB metric indicates the applicability of the tool in various domains. A score of 1 signifies that the system can be applied across multiple domains and targets, while a score of 0.5 indicates that the tool is specific to a particular domain or target. Furthermore, a score of 0.1 implies that the system is customized for a specific company. The rest of the parameters in VBM are defined in Table 2.

**Table 2.** Description of evaluation index for the business model.

| Components from Business Model Canvas | Criteria | Value Explanation |
|---|---|---|
| Key Partners | External partner reliability (EPR) | 1: none or small collaboration with external partners<br>0.5: external partners are necessary<br>0.1: mainly rely on external partners |
| Key Activities | Relative implementation difficulty (RID) | 1: easy to implement<br>0.5: can be design and implemented by in-house expertise<br>0.1: requires external expertise to design and implement |
| Key Resources | Resource accessibility (RA) | 1: existing resources<br>0.5: new but easy to reach<br>0.1: new and hard to reach |
| Value Propositions | Relative Benefits (RB) | 1: significant benefits in multiple areas |

| | | 0.5: significant benefits in one area, or minor benefits in multiple areas |
| | | 0.1: minor benefits in one area |
| Customer Relationships | New customer potential (NCP) | 1: easy to extend new customers |
| | | 0.5: hard to extend new customers |
| | | 0.1: aimed at keeping existing customers |
| Channels | Channel access difficulty (CAD) | 1: an existing channel |
| | | 0.5: new but easy to build |
| | | 0.1: new and hard to build |
| Customer Segments | Number of Segments (NS) | 1/(number of segments) |
| | | range: (0,1] |
| | | More segments may lead to more interest confliction. |
| Cost Structure | Relative Implementation Cost (RIC) | 1: high, and detailed economic feasibility analysis is required |
| | | 0.5: relatively average, and usually requires a first-order economic feasibility analysis |
| | | 0.1: limited to no cost involved to implement the opportunity |
| Revenue Streams | Familiarity and affordability (FA) | 1: familiar to customers and companies |
| | | 0.5: partly familiar to customers and companies |
| | | 0.1: totally new to customers and companies |

Based on the definition of each parameter and their value ranges, the range of VBM is (0, 64.8]. A large value represents a more significant commercial potential of the business model. The VBM value can be assigned to three different levels as Table 3 shows.

The low level denotes that the business model has little commercial value. The threshold 7.5 is calculated by setting TRL at 5, RB at 0.5, NS at 0.2, and all of the rest parameters at 0.5. An industrial digital solution with a TRL under 5 is still undergoing laboratory testing and validation, and real-world application tests have not yet been conducted. RB at 0.5 indicates that such a product has the potential to be sold to similar companies in the market. Moreover, more participating segments may lead to higher collaboration risks. For instance, it becomes challenging to reach a consensus where every segment involved is completely satisfied.

The medium level refers to a system that has limited commercial value in the market and several barriers to be applied in multiple areas, whereas the high level represents that the products have high commercial value in multiple areas. The threshold 21.9 is calculated by setting TRL at 7, RB at 1, NS at 1/3, and the rest of the parameters at 0.5. A product with a TRL above 7 is mature enough to be sold in the market. RB at 1 indicates that such a product has the potential to be sold to companies in different areas.

**Table 3.** Level of VBM values.

| Range of the VBM value | Level |
|---|---|
| (0, 7.5] | Low |
| (7.5, 21.9] | Medium |
| (21.9, 64.8] | High |

## 4 Case Studies

This research paper analyzes nine diverse instances of industrial digitalization, each one strategically implemented for the enhancement of energy efficiency and flexibility. These nine case studies are derived from nine distinct projects associated with the IEA-IETS Annex XVIII. This Annex is concerned with the integration of digitization, artificial intelligence, and other associated technologies to optimize energy efficiency and reduce greenhouse gas emissions within various industrial sectors [28].

A comprehensive analysis of all nine cases has been conducted using the business model canvas outlined in Fig. 1. This strategic management template is utilized to analyze each case's business model and understand their structure and strategies in detail. In addition to this, the commercial potential of each case is further assessed by computing their respective VBM values. This process provides valuable insights into the individual commercial viability and success potential of each case.

Furthermore, Table 4 provides an overview of the general information pertinent to the nine case studies. More detailed descriptions and findings for each case can be found in the referenced document [28].

**Table 4.** General information on the nine cases.

| No. | Company name | Sector | Topic |
|---|---|---|---|
| 1 | ArcelorMittal | Steel manufacturing | Scheduling a Galvanizing Line Using Ant-Colony Optimization to Minimize Costs and Energy Consumption |
| 2 | CanmetEN-ERGY/NRCan | Pulp and paper mill | Online Paper Machine Operation Optimization and Inefficiencies Diagnosis |
| 3 | Christonik | Transportation | Structured Data Collection and Analytics for Reduced Carbon Footprint Mobile |
| 4 | KMD A/S | General Production | Energy Key Production Insight |
| 5 | MayaHTT | Pulp and paper production | Pulp & Paper Production Quality Prediction |
| 6 | SDU-CEI | Horticulture | Digital Twin of Greenhouse Energy System |
| 7 | SDU-CEI | Horticulture | Digital Twin of Greenhouse Production Flow |
| 8 | Software AG | General Production | KI4ETA - Artificial Intelligence for Energy Technology and Applications in Production |
| 9 | Tata Steel B.V. | Steel manufacturing | SMEAT: Smart Heat Management |

# 5    Results

The nine cases are analyzed based on BMC in section 3.2, and the case information regarding the nine BMC blocks is summarized and compared in the following subsection. Their VBM scores are presented in this section as well.

## 5.1    Business Model Analysis

**Key Partners**. Table 5 illustrates the key partnerships in the selected cases, involving academic and industry stakeholders. These partnerships serve unique roles in developing and implementing industrial digital solutions, each contributing their distinct expertise. Academic partners are typically responsible for providing theoretical backing and contributing cutting-edge technology. On the other hand, industry partners supply necessary data and lend support in the testing and validation of digital tools.

Only a third of the cases involved academic partners, mainly due to the fact that the project leads hailed from academic institutions and were equipped to tackle theoretical challenges independently. Industrial partners are further divided into three categories: IT service providers, industry decision-makers and end-users, and investors. IT service providers support in establishing the IoT architecture, resolving hardware issues, and crafting software solutions. Industry decision-makers and end-users contribute real-world processing data and specific application requirements. Lastly, investors comprise of both investment companies and funding agencies.

**Table 5.** Summary and comparison of key partners in the nine cases.

| Key partners | Case 1 | Case 2 | Case 3 | Case 4 | Case 5 | Case 6 | Case 7 | Case 8 | Case 9 |
|---|---|---|---|---|---|---|---|---|---|
| Academic partners | ✓ | ✓ | | | | | | ✓ | |
| IT service providers | | ✓ | ✓ | | ✓ | | | ✓ | ✓ |
| Industry decision maker and end-user | | ✓ | | | ✓ | ✓ | ✓ | ✓ | ✓ |
| Investor | | | | ✓ | | | | | |

**Key Activities**. The key activities across the nine cases, as summarized in Table 6, demonstrate the range of technologies employed to create and deliver value in this study. These technologies include big data analytics, artificial intelligence (AI), machine learning (ML), Internet of Things (IoT), and digital twins (DT). Big data analytics methods are utilized to process raw industrial data, aimed at enhancing data quality and operational efficiency. ML and deep learning (DL) algorithms are employed as state-of-the-art AI technologies for prediction, decision-making, optimization, and recommendation across many domains. DT serves as a digital replication of a physical system, used to simulate the performance of the physical counterpart, leading to efficiency, productivity, and performance improvements. IoT pertains to a network of physical objects equipped with sensors and software that allow them to collect and exchange data over the internet.

**Table 6.** Summary and comparison of key activities in the nine cases.

| Key activities | Case 1 | Case 2 | Case 3 | Case 4 | Case 5 | Case 6 | Case 7 | Case 8 | Case 9 |
|---|---|---|---|---|---|---|---|---|---|
| Big data | | ✓ | ✓ | | ✓ | | | | |
| ML/DL | ✓ | ✓ | | ✓ | ✓ | | | ✓ | ✓ |
| DT | | ✓ | | | | ✓ | ✓ | | |
| IoT | | | | ✓ | | ✓ | | ✓ | |

**Key Resources**. Key resources include the vital assets and infrastructures required to deliver value propositions, such as skilled professionals, hardware/software facilities, and intellectual property rights. Within the context of industrial digitalization, the expertise of professionals and the software or platforms used for programming and simulation are considered critical resources.

**Value Propositions.** The value propositions of the nine cases are shown in Table 7. In the energy efficiency domain, the primary objectives are to reduce GHG emissions and minimize energy consumption. This involves implementing technologies that enhance energy resource utilization, such as equipment upgrades and energy-saving practices. By reducing energy waste and improving overall efficiency, the purpose is to mitigate the environmental impact associated with energy generation and consumption.

In the energy flexibility domain, the focus shifts toward identifying and implementing optimal practices in industrial processing. The goal is to optimize processes efficient energy resource utilization, leading to reduced energy costs and production time. While reducing energy costs and production time are key targets in the energy flexibility domain, it's important to consider other value propositions as well. Operational costs play a significant role in determining industrial best practices, so any improvements in energy efficiency or flexibility should aim to minimize these costs. Additionally, the production quality is crucial in ensuring the reliability and performance of industrial systems, and it should be considered alongside other factors when designing and implementing industrial digital solutions for energy efficiency and flexibility in industry.

**Table 7.** Summary and comparison of value propositions in the nine cases.

| Value propositions | Case 1 | Case 2 | Case 3 | Case 4 | Case 5 | Case 6 | Case 7 | Case 8 | Case 9 |
|---|---|---|---|---|---|---|---|---|---|
| Improve energy efficiency | | ✓ | ✓ | ✓ | | | ✓ | ✓ | ✓ |
| Reduce GHG emissions | | ✓ | ✓ | ✓ | ✓ | ✓ | | | ✓ |

| | Case 1 | Case 2 | Case 3 | Case 4 | Case 5 | Case 6 | Case 7 | Case 8 | Case 9 |
|---|---|---|---|---|---|---|---|---|---|
| Reduce energy cost | | | | ✓ | ✓ | | | | |
| Reduce production time | | | | ✓ | | | | | |
| Optimize process | ✓ | | | ✓ | ✓ | ✓ | ✓ | | |
| Improve production quality | ✓ | | | | | | ✓ | | |

**Customer Relationships**. The term 'customer relationships' denotes the interaction and services exchanged between system providers and customers. In the context of energy efficiency and flexibility, industrial digital solutions are customized to meet unique demands. Consequently, technology providers are expected to offer timely consultation and regular maintenance services.

**Channels**. To facilitate smooth customer interaction, service providers must establish a stable and easily accessible communication channel. This could be through online platforms or telephonic services, enabling efficient and convenient exchanges.

**Customer Segments**. Table 8 summarizes the customer segments in the cases. These groups are categorized based on their roles in the project. This study focuses on the technical segments of digitalization system implementation, while excluding the management and financial departments from its scope. Initially, the IT department responsible for production data management should be involved to provide top-tier data, thereby ensuring accurate analysis and informed decision-making. Typically, this data collection and management process is led by the technology departments. Subsequently, IT departments also infuse their industry-specific knowledge into the project, offering valuable insights on technical aspects, standards, regulations, and best practices related to the industrial digital solutions. Finally, the Research and Development (R&D) department illuminates the challenges and optimization necessities within the production workflow, leveraging their expertise to identify inefficiencies, bottlenecks, and areas for improvement.

**Table 8.** Summary and comparison of customer segments in the nine cases.

| Customer segments | Case 1 | Case 2 | Case 3 | Case 4 | Case 5 | Case 6 | Case 7 | Case 8 | Case 9 |
|---|---|---|---|---|---|---|---|---|---|
| R&D Dept. | ✓ | ✓ | | | | | | | ✓ |
| Energy Dept. | | ✓ | | | | | | ✓ | ✓ |
| IT Dept. Technology | ✓ | | ✓ | ✓ | ✓ | ✓ | ✓ | | |

**Cost Structure**. Table 9 demonstrates the cost structure of the cases. It's crucial to identify all significant costs and expenses tied to the operation of your business, including fixed and variable costs, economies of scale, and resource allocation. When managing a business for digitalization systems, it is essential to consider both fixed and variable costs. The fixed costs encompass expert hours for system development and the licensing costs of software and platforms. On the other hand, variable costs relate to devices and expert hours required for system updates and maintenance.

**Table 9.** Summary and comparison of the cost structure in the nine cases.

| Cost structure | Case 1 | Case 2 | Case 3 | Case 4 | Case 5 | Case 6 | Case 7 | Case 8 | Case 9 |
|---|---|---|---|---|---|---|---|---|---|
| Expert hours | ✓ | ✓ | ✓ | ✓ | ✓ | ✓ | ✓ | ✓ | ✓ |
| License | | ✓ | ✓ | | ✓ | ✓ | ✓ | ✓ | ✓ |
| System update/maintenance | ✓ | ✓ | ✓ | ✓ | ✓ | ✓ | ✓ | ✓ | ✓ |

**Revenue Streams**. Within the context of this study, revenue streams are defined as the various modes of financial inflow. These streams epitomize the agreed upon forms of remuneration, carefully negotiated and established between system providers and customers. These arrangements not only underscore the business' profitability but also reflect the value proposition offered to the consumers.

## 5.2    Value of Business Model Evaluation

The data illustrated in Table 10 reveals that six out of the nine industrial digital solutions occupy a medium position, signifying that a majority of the industrial digital solutions hold commercial potential, but require additional verification. Primarily, in most instances, the Technology Readiness Levels (TRLs) of these systems are classified at level 6. This indicates that these industrial digital solutions have undergone successful validation and demonstration in a context that is relevant to their designed functions. However, to enhance the TRLs, it's necessary to undertake further testing in a real-world operational environment.

Moreover, although these industrial digital solutions bring significant advantages in certain areas, their applicability in different domains tends to be restricted, leading to a Relevance-Breadth (RB) score of 0.5. To illustrate, the software deployed in the third case study has successfully enhanced energy efficiency by up to 20%. However, its ability to be adapted to optimize different objectives, such as operational cost, or its applicability in other industrial sectors, might be constrained. Therefore, its VBM level is medium as shown in Table 10.

**Table 10.** Evaluation scores of parameters in VBM

| Cases | TRL | RB | EPR | RID | RA | NCP | CAD | NS | FA | RIC | VBM | Level |
|---|---|---|---|---|---|---|---|---|---|---|---|---|
| 1 | 9 | 1 | 1 | 0.5 | 0.1 | 1 | 0.5 | 0.2 | 0.5 | 0.1 | **36** | **High** |
| 2 | 6 | 0.5 | 0.5 | 0.5 | 0.1 | 0.1 | 0.5 | 1 | 0.5 | 0.1 | **10.2** | **Medium** |
| 3 | 8 | 0.5 | 0.1 | 1 | 0.5 | 1 | 0.5 | 1 | 0.5 | 0.1 | **19.2** | **Medium** |
| 4 | 6 | 0.5 | 1 | 0.1 | 0.5 | 0.1 | 0.5 | 1 | 0.5 | 0.5 | **10.5** | **Medium** |

| 5 | 7 | 1 | 0.5 | 1 | 0.1 | 0.5 | 0.5 | 1 | 0.5 | 0.5 | **27.3** | **High** |
| 6 | 6 | 0.5 | 0.5 | 1 | 1 | 0.1 | 0.5 | 1 | 0.5 | 0.5 | **13.2** | **Medium** |
| 7 | 6 | 0.5 | 0.5 | 1 | 1 | 0.1 | 0.5 | 0.5 | 0.5 | 0.1 | **12.9** | **Medium** |
| 8 | 6 | 1 | 0.1 | 0.5 | 0.1 | 0.5 | 0.5 | 1 | 0.5 | 0.5 | **18** | **Medium** |
| 9 | 3 | 0.1 | 0.5 | 0.5 | 0.5 | 0.1 | 0.5 | 0.33 | 0.5 | 0.1 | **0.94** | **Low** |

## 6 Discussion

### 6.1 Strengths and Weaknesses

The nine cases in this paper are all specifically designed with a focus on enhancing energy efficiency and flexibility within various industries. Given the European Union's ambitious objective of a 55% reduction in GHG emissions by 2030 [29], there is a significant potential for growth in the market for industrial digitalization tools geared towards decreasing energy consumption and GHG emissions.

Current industrial digital solutions have demonstrated their proficiency in offering solutions for improving energy efficiency and reducing operational costs. For example, the digitalization framework in case 2 helps the pulp and paper mill company save about 11% energy. This is achieved through the deployment of advanced digitalization methodologies including Machine Learning/Deep Learning (ML/DL), Digital Twins (DT), and the Internet of Things (IoT). Additionally, these industrial digital solutions often offer customization to suit specific partners and objectives, thus facilitating flexible requirements from the client's end. Moreover, the development cost of these digital solutions is relatively manageable, usually entailing expenses such as expert personnel salaries, licensing fees for specific software or platforms, and certain hardware devices.

Despite these strengths, there are limitations that must be acknowledged. Firstly, the discrete scale used in the VBM metric is derived from [10, 12], which might not be the optimal scale for evaluation as how close to a specific value is not defined. Secondly, except for cases 1 and 3, the remaining digital solutions are still in developmental phases, not yet primed for commercial applications. They necessitate extensive real-world testing to ensure reliability and effectiveness. In addition, digital systems also contribute to GHG emissions through electricity consumption by servers, computing equipment, and cooling systems. Achieving net energy savings in both the industry system and digitalization tools is therefore paramount. Moreover, the discussed industrial digital solutions were sector-specific, limiting their applicability. Therefore, tech providers should broaden product adaptability across sectors, potentially expanding their customer base, supporting commercial success, and advancing energy efficiency and GHG reduction goals.

### 6.2 Relationships Between BMC Components

The results show that there are differences in key partners key activities, customer segments from value proposition perspective as shown in Table 11, but not in the other BMC components. Commonly, timely information exchange with customer departments is crucial for discussing requirements and maintenance services in the development of industrial digital solutions. Therefore, it is essential to establish a stable and accessible communication channel. Furthermore, careful calculation and discussion of costs and budget between the system provider and customer, along with negotiation of financial inflow, are critical for ensuring successful system development. Moreover, both industrial and academic partners can contribute to the big-data analysis and AI frameworks in distinct aspects. Such methods are leveraged in some software or hardware platforms by skilled professionals, which are the key resources.

**Table 11.** Relationships between BMC components

| Value propositions | Improve energy efficiency | Reduce GHG emissions | Optimize process | Improve production quality | Reduce energy cost | Reduce production time |
|---|---|---|---|---|---|---|
| **Key partners** | Industry decision maker and end-user; IT service providers | | | | | |
| | Academic partners | | | | | |
| | Investor | | | | | |
| **Key activities** | Big data; Machine learning; IoT | | | | | |
| | DT | | | | DT | |
| **Customer segments** | IT Department | | | | | |
| | R&D Department; Energy Department | | | | | |

## 6.3 Recommendations on digital solution development for enabling energy efficiency and flexibility in industry

This paper presents an improved and comprehensive set of recommendations designed to enhance the effectiveness and value proposition of industrial digital solutions, with a particular focus on big data-driven tools and services:

- Fostering cross-sector collaboration: Cross-sector collaboration is a pivotal strategy for developing digitalization solutions [30]. Industry stakeholders should emphasize the collection and integration of high-quality, multi-sourced data, as well as the use of advanced data analytics technologies in their operations. This requires forging partnerships with IT companies offering AI-integrated, data-driven products and services. Additionally, collaboration with research institutes can offer access to state-of-the-art solutions, fostering knowledge sharing and further enhancing data utilization.
- Prioritizing comprehensive testing and validation: The operational effectiveness of industrial digital solutions relies heavily on extensive testing and validation in real-world environments [31]. Such practices ensure the systems can reliably handle real-life situations, identify potential bottlenecks or weaknesses before full-scale deployment. This can lead to a more secure, reliable, and efficient system.
- Extending value propositions: The value proposition of digital solutions should not be solely technology-focused [32]; it should also include aspects like operational costs and production quality. In a competitive industrial environment, organizations should strive to enhance efficiency and product quality, two key performance indicators directly influenced by digitalization.

- Enhancing product adaptability: Digital solution providers should focus on improving the adaptability of their systems for wider application scenarios [33]. This broadens the customer base by catering to diverse industry requirements, which subsequently enhances the commercial viability and impact of the solutions.

- Providing user-friendly platforms: From the customer's perspective, it's vital to have a user-friendly platform that provides accurate analyses [34]. The ease of use and precision of insights significantly influence customer satisfaction, facilitating more informed decision-making and promoting adoption of the solution.

- Adopting transparency: Service providers should offer clear and understandable recommendations to clients [35]. Hence, the application of explainable AI technologies is highly recommended in these systems. Such transparency fosters trust between providers and users, making it easier for clients to understand the value of the solution and how it contributes to their operations.

## 7    Conclusion

This study conducted an in-depth analysis and provided a critical discussion surrounding digital solutions designed for energy efficiency and flexibility within the industrial sector. Primary data on nine pertinent cases was meticulously compiled via survey methods, which were subsequently evaluated and compared through the lens of the Business Model Canvas framework. In a novel approach, an evaluative index known as the Value of Business Model (VBM) was introduced with the purpose of assessing the business viability and potential of these digital solutions.

Our findings underscore the necessity for extensive real-world testing and validation to elevate the technical robustness of these digital solutions. Additionally, to optimize energy efficiency and flexibility in industry, operational costs and product quality should be elevated to the status of key metrics, alongside traditional considerations of energy consumption and GHG emissions. It is equally vital that the digital solutions possess a degree of extensibility to enable their adoption across wider scenarios, consequently broadening the potential customer base. A system's inherent flexibility and scalability, therefore, are paramount attributes that enable adaptation to diverse needs and accommodate future expansion.

In summary, it is of utmost importance to orchestrate symbiotic, intelligent digital solutions capable of informing and guiding decisions concerning energy efficiency and energy flexibility. Drawing on the insights gleaned from the case studies featured in this study, future research endeavors can further refine and enhance industrial digital solutions. It is recommended to design digital solutions that are user-friendly, transparent, and possesses the ability to maximize potential for application across a range of contexts. This approach will not only improve usability but also foster a wider acceptance and implementation of digital solutions in various industrial sectors.

**Acknowledgements.** This paper is part of the IEA IETS Task XVIII: Digitalization, Artificial Intelligence and Related Technologies for Energy Efficiency and GHG Emissions Reduction in Industry, funded by the Danish funding agency, the Danish Energy Technology Development and Demonstration (EUPD) program, Denmark (Case no.134-21010), and the project "Data-driven best-practice for energy-efficient operation of industrial processes - A system integration approach to reduce the CO2 emissions of industrial processes" funded by EUDP (Case no.64020-2108).